\documentclass[aps,prb,preprint,showpacs,preprintnumbers,amsmath,amssymb]{revtex4}

\usepackage{graphicx}

\begin{document}

\title{Exchange interaction and stability diagram of coupled quantum dots in magnetic fields}

\author{L.-X. Zhang, D. V. Melnikov, and J.-P. Leburton}
\affiliation{
Beckman Institute for Advanced Science \& Technology and Department of Electrical and Computer Engineering,
University of Illinois at Urbana-Champaign, Urbana, Illinois 61801
}

\date{\today}

\begin{abstract}
The charge stability diagram for two coupled quantum dots containing up to two electrons is computed in magnetic fields. One- and two-particle Schr\"odinger equations are solved by exact diagonalization to obtain the chemical potentials and exchange energy in these systems. By analyzing the chemical potentials variation with external biases and magnetic fields, it is possible to distinguish between the weak and strong inter-dot couplings. The variation of the chemical potential curvatures and the double-triple point separations in the stability diagrams confirms the inter-dot coupling decrease with increasing magnetic fields. The computed exchange energies are also found to be significantly smaller than the values estimated from the stability diagram.
\end{abstract}

\pacs{73.21.La, 73.21.-b}

\maketitle

Coupled quantum dots (QDs) are of particular importance for spin-based quantum computation because universal quantum logical gates (such as a Control-NOT gate) can be realized via the interaction between two quantum bits (qubits), {\it i.e.}, the spins of two electrons, each trapped in one quantum dot.\cite{Loss} In such devices, the interaction between the two spins is proportional to the exchange energy $J$, which is equivalent to the splitting between the lowest singlet and triplet two-electron states. 

While extensive theoretical work focuses on the dependence of $J$ on the system paramenters such as the inter-dot separation, the tunneling barrier between the QDs, and the external magnetic field,\cite{Burkard,Hu,Harju,Szafran} the charge stability diagram of coupled QDs \cite{Wiel} has been studied to a lesser extent. Meanwhile, recent advances in experimental techniques have made it possible to study coupled QDs in the few-electron regime when each QD contains only one conduction electron (see, {\it e.g.,} Refs. [\onlinecite{Wiel, exp, Hatano}]). In this case the stability diagram becomes a powerful tool to study inter-dot coupling and electronic transport through double QD systems. Analysis of the stability diagram and its evolution in magnetic fields allows one to estimate the values of the exchange energy as was demonstrated recently in the case of the two laterally coupled vertical QDs.\cite{Hatano}

In general, in the stability diagram the boundaries between distinct stable charge states, {\it i.e.}, between the states with fixed number of electrons $N_1$ and $N_2$ in each of the coupled dots, are represented as functions of the two controlling gate biases, one for each dot.\cite{Wiel} These equilibrium charges are determined from the condition that the chemical potential of the QD structure $\mu(N_1+N_2)$ defined as: \cite{Wiel}
\begin{equation}
\mu(N_1+N_2) = E_G(N_1+N_2) - E_G(N_1+N_2-1),
\label{eqn:mu}
\end{equation}
where $E_G(N)$ is the ground state energy of the $N$-electron state, is less than that of the leads (source and drain).

In this paper, we numerically compute the stability diagram in coupled QDs with $N_1+N_2\leq 2$ 
electrons in external magnetic fields, and investigate its properties for different inter-dot coupling strengths. The Hamiltonian for the coupled system is given by
\begin{equation}
H({\bf r_1},{\bf r_2}) = H_{orb}+H_Z,
\label{eqn:H_tot}
\end{equation} 

\begin{equation}
H_{orb}=h({\bf r_1})+h({\bf r_2})+C({\bf r_1},{\bf r_2})
\label{eqn:H_orb}
\end{equation} 

\begin{equation}
h({\bf r}) = \frac{1}{2m^*}({\bf p}+\frac{e}{c}{\bf A})^2 + V({\bf r}),
\label{eqn:H_single}
\end{equation} 

\begin{equation}
C({\bf r_1},{\bf r_2})=e^2/\epsilon|{\bf r_1}-{\bf r_2}|
\label{eqn:C}
\end{equation} 

\begin{equation}
H_Z = g\mu_B\sum_{i}{\bf B}\cdot{\bf S_i}
\label{eqn:H_Z}
\end{equation}

\noindent Here, $m^*=0.067m_e$ is the electron effective mass, $\epsilon=13.1$ is the dielectric constant, 
$g=-0.44$ is the g-factor in GaAs, $\mu_B$ is the Bohr magneton and ${\bf A} = \frac{1}{2}[-By,Bx,0]$ is the vector potential for the constant magnetic field $B$ oriented perpendicular to the QD plane ($xy$-plane). The Zeeman effect is included in all our calculations except otherwise mentioned.

The confinement in double QDs is simulated by the following model potential:\cite{Hu, Szafran}
\begin{equation}
V({\bf r}) = -V_Le^{-[(x+d/2)^2+y^2]/R^2}-V_Re^{-[(x-d/2)^2+y^2]/R^2},
\label{eqn:potential}
\end{equation}
where $V_L$ and $V_R$ are the depth of the left and right dots (equivalent to the QD gate voltages in experimental structures \cite{Wiel,exp,Hatano}) which can be independently varied, $d$ is the inter-dot separation, and $R$ is the radius of
the dot ($R=30$ nm).

The single-particle Hamiltonian Eq. (\ref{eqn:H_single}) and the orbital part of the two-electron Hamiltonian Eq. (\ref{eqn:H_orb}) are diagonalized numerically to obtain the ground state eigenvalues which are used to compute the chemical potentials $\mu(1)$ and $\mu^S(2)$ [$\mu^T(2)$] according to Eq. (\ref{eqn:mu}) [note that $E_G(0)\equiv 0$] and the exchange energy $J$ as
\begin{equation}
J = E_G^T(2) - E_G^S(2).
\label{eqn:J}
\end{equation}
Above, the symbols S and T pertain to the singlet state and the lowest triplet state, respectively.

\begin{figure}[htp]
\includegraphics[width=6.5cm]{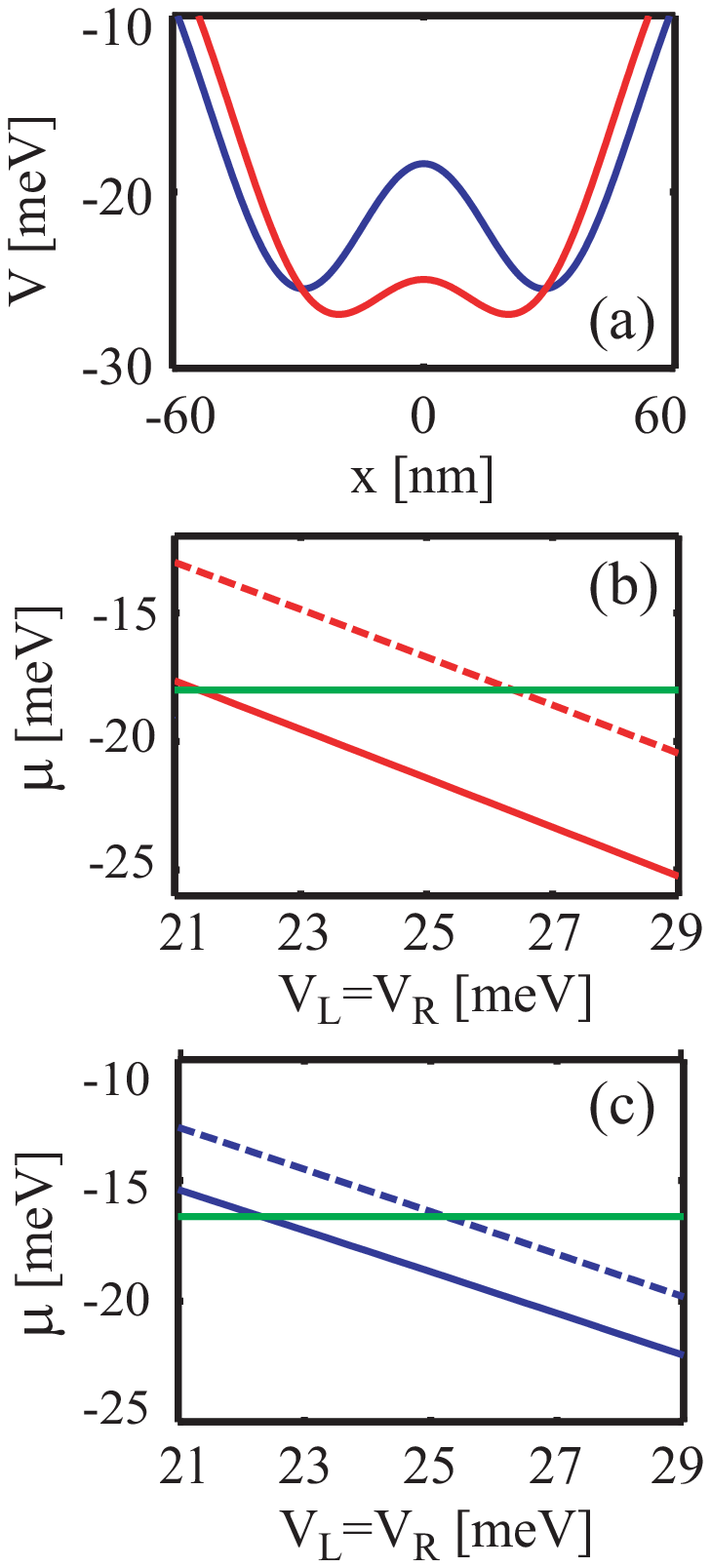}
\caption{\label{fig:fig1}
(Color online) 
(a) The confinement potential for $d = 50$ nm (red) and $d = 60$ nm (blue) at $V_L = V_R = 25$ meV. Chemical potentials $\mu(1)$ (solid) and $\mu^S(2)$ (dashed) vs. $V_L=V_R$ for (b) $d = 50$ nm and (c) $d = 60$ nm. In both of them the horizontal line indicates the values of the chemical potential at which the contours in Fig. \ref{fig:fig3} are drawn.
}
\end{figure}

We consider two coupled QD systems with different inter-dot separations $d = 50$ nm (strong coupling) and $d = 60$ nm (weak coupling). The confinement potential for the two cases is plotted in Fig. \ref{fig:fig1}(a) where it is seen that aside from the increased lateral separation the barrier is also higher in the latter case (1.98 vs. 7.10 meV). Fig. \ref{fig:fig1}(b) and (c) show the chemical potentials $\mu(1)$ and $\mu^S(2)$ which in the absence of the magnetic field decrease almost linearly as $V_L$ and $V_R$ simultaneously increase. The quasi-linear decrease of the chemical potentials with slightly larger slopes than shown was found for various magnetic fields as well.\cite{Zhang}

\begin{figure}[htp]
\includegraphics[width=10cm]{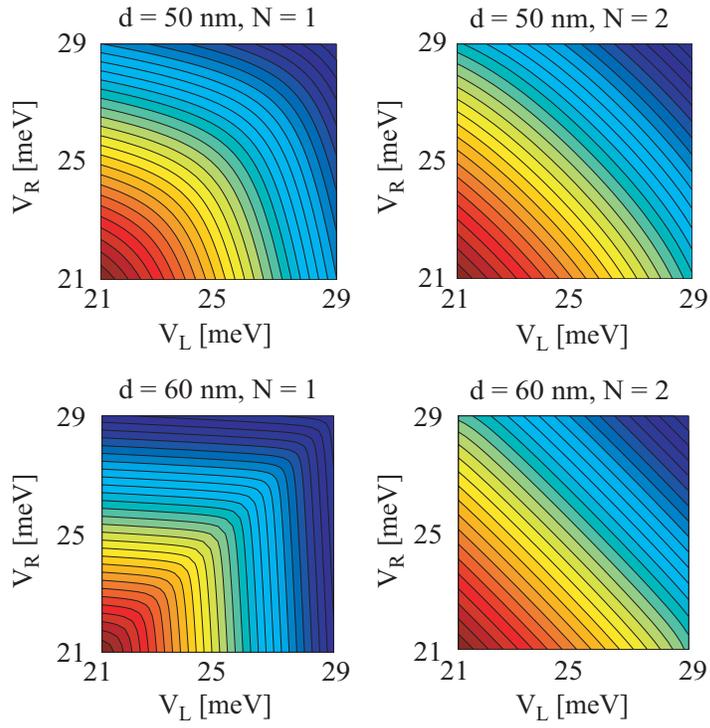}
\caption{\label{fig:fig2}
(Color online) 
Surface (contour) plots of the total energies for $N=1$ (left column) and $N=2$, singlet (right column) at $d=50$ nm (top row) and $d=60$ nm (bottom row). In all plots the energies decrease from the lower left to the upper right corner.}
\end{figure}

We first analyze the dependences of the total energies $E_G(1)$ and $E_G^S(2)$ for one and two electrons on $V_L$ and $V_R$ shown in Fig. \ref{fig:fig2}. We see that in the one-electron case (left column), the curvature of $E_G(1)$ in the region where $V_L$ and $V_R$ are near each other ($V_L\sim V_R$) is larger for $d=60$ nm than for $d=50$ nm because of the weaker coupling between the QDs in the former case. We also note that for both values of the inter-dot distance, the curvatures of the energy contour plots increase along the main diagonal since the coupling between the dots decreases as $V_L=V_R$ gets larger.

However, when the two electrons populate the QD system, the situation becomes radically different: in the weak coupling case ($d=60$ nm, bottom right), the total energy of the two electron system in the $V_L\sim V_R$ region is almost linearly dependent on $V_L$ ($V_R$), {\it i.e.}, the curvature is vanishingly small, while for $d=50$ nm (top right) the energy curves clearly exhibit a non-linear behavior with non-zero curvatures. The large overlap between the electrons in the strong inter-dot coupling case ($d=50$ nm) is responsible for the smooth non-linear dependence of the energy on $V_L$ ($V_R$). However, in the weak-coupling case ($d=60$ nm), the two electrons are well localized in the individual QDs by Coulomb repulsion and the large barrier between the dots, so that the potential change in one QD caused by the variation of $V_L$ ($V_R$) does not affect the electron charge distribution but only acts as a constant addition to the total energy. This leads to a linear dependence of the total energy on $V_L$ ($V_R$). \cite{note5} When the difference between $V_L$ and $V_R$ becomes sufficiently large to overcome the Coulomb repulsion, the two electrons move into one QD. This is accompanied by a change in the slope of the energy curves which become either horizontal or vertical as $V_L$ (or $V_R$) no longer affects the total energy and which corresponds to the (0,2)/(2,0) regions on the stability diagram (not shown). \cite{Wiel} We emphasize that the observed quasi-linear behavior of the total energy $E_G(2)$ when $V_L\sim V_R$ in the weak coupling regime ($d=60$ nm) is physically different from the situation in two coalesced dots where both $E_G(1)$ and $E_G(2)$ are also straight lines perpendicular to the main diagonal in the $V_L - V_R$ plane.\cite{Wiel} This is because in that case one deals with a single QD and changing $V_L$ ($V_R$) modifies the total energy of the system.

\begin{figure}[htp]
\includegraphics[width=10cm]{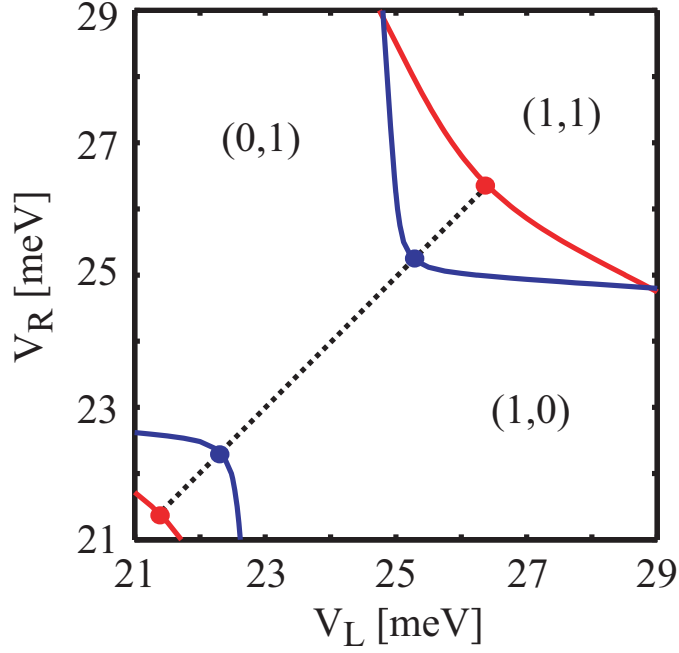}
\caption{\label{fig:fig3}
(Color online) 
Contour plots of the chemical potentials $\mu(1)$ and $\mu^S(2)$ as functions of $V_L$ and $V_R$ for $d = 50$ nm (red) and $d = 60$ nm (blue). The turning points on the contour lines are indicated by solid dots and the dotted line is a guide for the eyes along the main diagonal ($V_L=V_R$). The numbers on the left (right) within parentheses give the electron number in the left (right) dot [the (0,0) region is located at the lower left corner below the $\mu(1)$ branch for $d = 50$ nm].}
\end{figure}

Fig. \ref{fig:fig3} displays the contour plots of $\mu(1)$ (lower branches) and $\mu^S(2)$ (upper branches) as functions of $V_L$ and $V_R$ at zero magnetic field. We choose constant values of $\mu(1) = \mu^S(2)=-18$ meV for $d = 50$ nm and $\mu(1) = \mu^S(2)=-16.5$ meV for $d = 60$ nm as the reference values of the chemical potential in the source/drain of the QD device.\cite{note4} In       Fig. \ref{fig:fig3} we can recognize four regions corresponding to four stable charge states with different numbers of electrons in each dot [the numbers in the parentheses in each region give the number of electrons in the (left, right) QD] separated by the chemical potential contours and the main diagonal $V_L=V_R$. At the turning point on each branch along the main diagonal, three stable charge states coincide in terms of the total energy of the system. The distance between the turning points is the so-called double-triple point (DTP) separation (also called the anti-crossing separation).\cite{Wiel, Hatano} From Fig. \ref{fig:fig3} we also observe that the DTP separation $\Delta V_L=\Delta V_R=5.00$ meV in the $d=50$ nm case is significantly larger than the corresponding value $\Delta V_L=\Delta V_R=2.93$ meV in the $d=60$ nm case. Furthermore, the curvature of the branches around the DTP is smaller for $d=50$ nm than for $d=60$ nm. According to the ``classical" theory,\cite{Wiel} a smaller DTP separation (or equivalently a larger curvature of the chemical potential contour lines) indicates a weaker inter-dot coupling which is consistent with our findings.

From the data in Table \ref{tab:tab1}, we note that for $d=50$ nm, the curvature (magnitude) $\kappa(2)$ of the $\mu(2)$ curve is smaller than the curvature (magnitude) $\kappa(1)$ for $\mu(1)$, while in the $d=60$ nm case $\kappa(1)<\kappa(2)$. This peculiar behavior can be clarified by noting that both $\kappa(1)$ and $\kappa(2)$ are determined by the differences between the corresponding curvatures of the total energy whose behavior in the voltage plane is discussed above. This indicates that in general, all being equal, in the weak-coupling regime the curvature of the chemical potential for two electrons is larger than that one for one electron, $\kappa(2)>\kappa(1)$, while in the strong coupling regime, the opposite relationship $\kappa(2)<\kappa(1)$ holds.

\begin{figure}[htp]
\includegraphics[width=10cm]{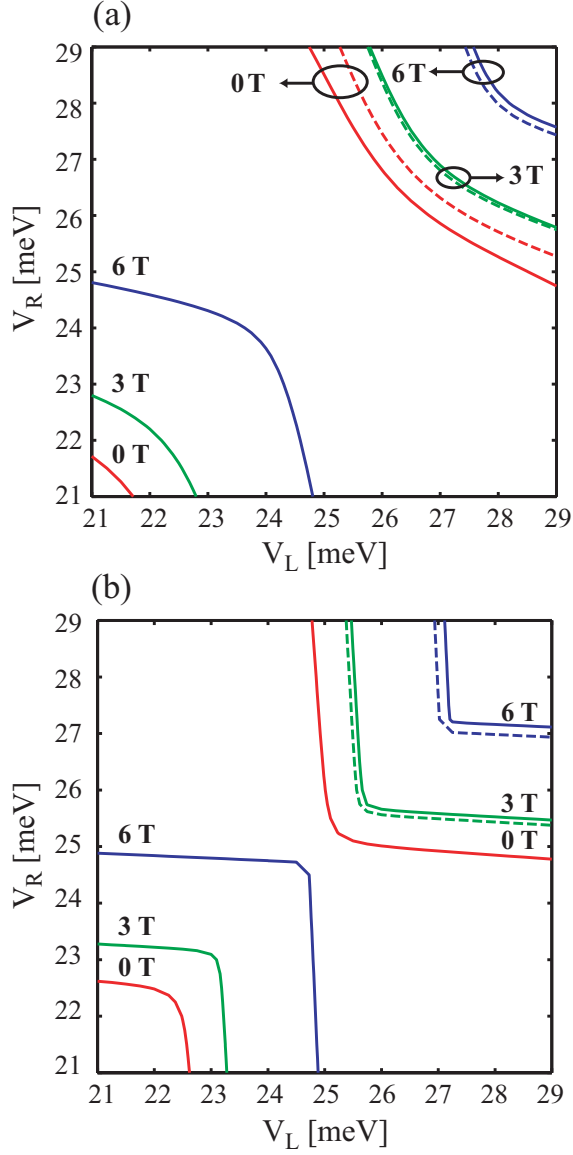}
\caption{\label{fig:fig4}
(Color online) (a) Contour plots of the chemical potentials $\mu(1)$ (single electron state, lower branches, solid curves), $\mu^S(2)$ (two-electron singlet state, upper branches, solid curves), and $\mu^T(2)$ (the lowest two-electron triplet state, upper branches, dashed curves) at different magnetic fields for $d=50$ nm. (b) Same as (a) but at $d=60$ nm. In the case of $d=60$ nm and $B=0$, the contour lines for $\mu^S(2)$ and $\mu^T(2)$ are indistinguishable on the scale of the figure.}
\end{figure}

\begin{figure}[htp]
\includegraphics[width=10cm]{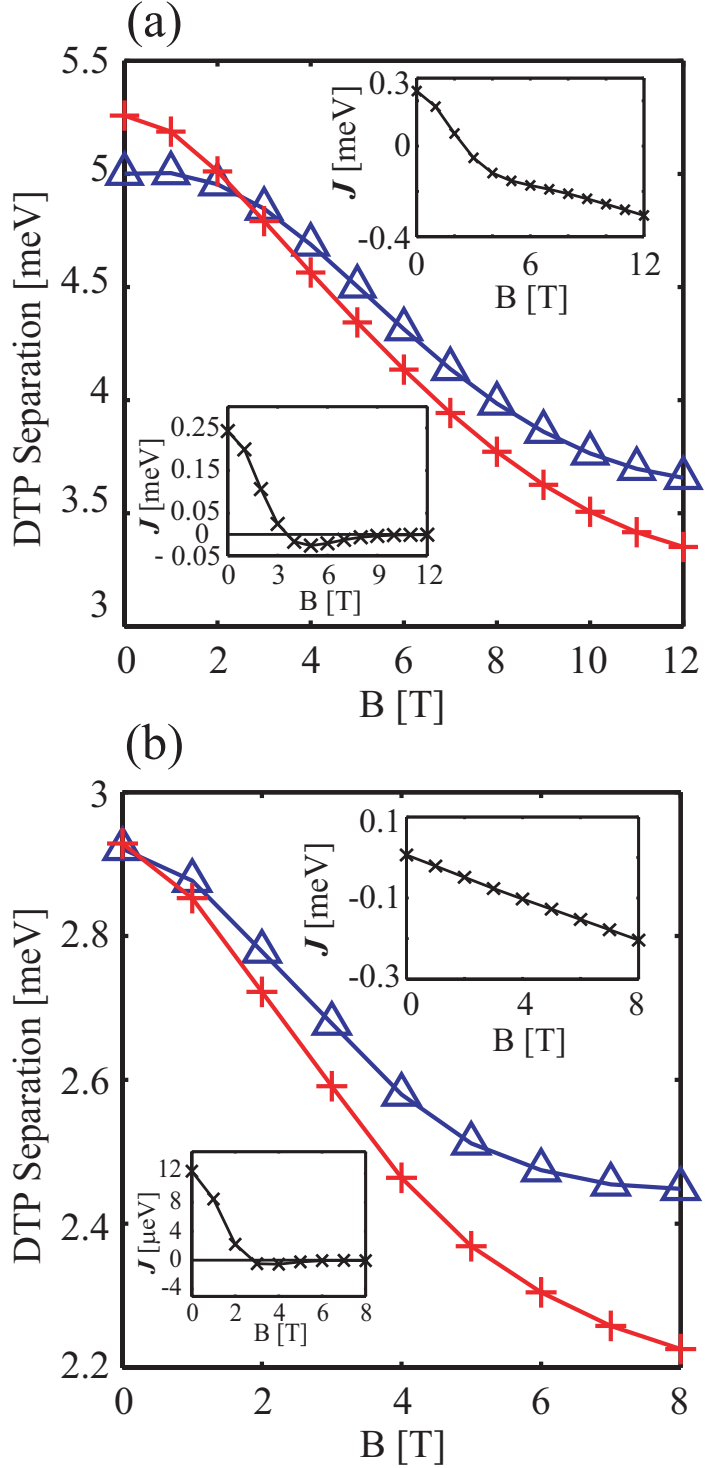}
\caption{\label{fig:fig5}
(Color online) Double-triplet point separation along $V_L$ (or $V_R$, $V_L=V_R$) axis as a function of the magnetic field $B$ for (a) $d = 50$ nm and (b) $d = 60$ nm. The data for singlet and the lowest triplet states are labeled by "$\Delta$" (blue curve) and "+" (red curve), respectively. The upper (lower) inset in each figure shows the exchange energy $J$ as a function of the magnetic field with (without) the Zeeman effect. The data in the insets are obtained at $V_L=V_R=25$ meV.}
\end{figure}

In the presence of the magnetic field, the curvatures of the chemical potential contours also increase as can be seen in Fig. \ref{fig:fig4}(a) and (b) where we again plot the chemical potential contours for $\mu(1)$, $\mu^S(2)$ and $\mu^T(2)$ at constant reference values of $\mu(1)=\mu^S(2)=\mu^T(2)= -18$ ($-16.5$) meV for $d=50$ (60) nm at $B=0$, 3 and 6 T. Note the order of the contours for $\mu^S(2)$ and $\mu^T(2)$ at different magnetic fields. As the magnetic field increases, the contours shift from the lower left corner to the upper right corner because the single-particle eigenenergies increase.\cite{Harju} In addition to the curvature increase, the DTP separation becomes smaller at larger magnetic field for both singlet and the lowest triplet states [for the detailed explanation of this effect, see the discussion on Fig. \ref{fig:fig5}(a)]. From the changes in the curvature and DTP separation, one concludes that the magnetic field indeed induces a quantum mechanical decoupling between the two dots and results in magnetic localization of electrons in each dot. By comparing the chemical potential curves in Fig. \ref{fig:fig4}(a) with the corresponding ones in Fig. \ref{fig:fig4}(b), we see that in the latter case the chemical potential contours have much larger curvatures than in the former case (see Table \ref{tab:tab1} for details) due to the increased inter-dot decoupling and for each value of the magnetic field the DTP separation for $d=60$ nm is more than 60\% smaller than for $d=50$ nm.

From Table \ref{tab:tab1}, it is also shown that the curvatures $\kappa(1)$ and $\kappa(2)$ progressively increase as the magnetic field becomes larger. This is due to enhanced localization of electrons caused by the magnetic field. The magnetic localization in the weak coupling case became prevalent at lower fields  than in the strong coupling situation [see lower insets of Fig. \ref{fig:fig5}], which is manifested by a more rapid increase in the curvature of chemical potential contours.

Figures \ref{fig:fig5}(a) and (b) show the extracted DTP separation along $V_L$ (or $V_R$, $V_L=V_R$) axis as a function of magnetic fields for $d=50$ nm and $60$ nm inter-dot separations, respectively. In each plot the data are shown for the singlet and lowest triplet states. Note that at $B=0$ the DTP separation for the singlet state is smaller than that for the lowest triplet state because the singlet is the ground state, while at larger $B$ fields, the lowest triplet state becomes the ground state and the order of the DTP separations is reversed. In both (a) and (b), the DTP separation for the lowest triplet state decreases faster with $B$ fields than that for the singlet state. This is because the DTP separation is proportional to $\mu(2)-\mu(1)=E_G(2)-2E_G(1)$ for a fixed $V_L=V_R$ on the main diagonal of the stability diagram (see Figs. \ref{fig:fig1} and \ref{fig:fig3}). For the singlet state, $E_G(2)$ does not change with the $B$ field while $E_G(1)$ decreases with the $B$ field due to the Zeeman effect, therefore the Zeeman contribution to $\mu(2)-\mu(1)$ increases with the $B$ field. For the triplet state, the Zeeman contributions to $E_G(2)$ and $2E_G(1)$ cancel out, and $\mu(2)-\mu(1)$ is not affected by the $B$ field. The decrease of the DTP separation in the magnetic field was also recently observed experimentally.\cite{Hatano} The upper (lower) inset in each figure shows the corresponding exchange energy $J$ as a function of the magnetic field calculated by Eq. (\ref{eqn:J}) with (without) the Zeeman effect.
In both cases, the Zeeman effect induces a linear depenedence of $J$ on $B$. However, in (a) given the strong coupling between the dots, the orbital contribution to $J$ dominates at low $B$ fields before being overcome by the Zeeman induced decrease at higher field; in (b), $J$ is totally dominated by the Zeeman contribution, which decreases linearly with the $B$ field. Comparison of the $B$ field dependences of the DTP separation and exchange energy in the absence of the Zeeman effect in Fig. \ref{fig:fig5} shows that the latter saturates at much lower values of the magnetic field than the former. This is because the DTP separation is determined by the Coulomb interaction between electrons which decreases as the electrons become localized by the magnetic field in individual dots (within the Heitler-London approximation, this decrease is proportional to $B^{-2}$, Ref. \onlinecite{dmm31}), while the exchange energy in absence of the Zeeman effect approaches zero much faster than the Coulomb interaction since it is proportional to the overlap between the individual electron wave functions that decays exponentially fast in strong magnetic fields.\cite{Burkard, dmm31} 

It is also interesting to compare the exact values of the exchange energy (see the insets in Fig. \ref{fig:fig5}) with those extracted from the stability diagrams in magnetic fields using the Hubbard model.\cite{Burkard, Hatano} According to this model,
$J_{est}=4t^2/(V_{intra}-V_{inter})$ where $2t$ is the tunnel (symmetric-asymetric) splitting, $V_{intra}$ and $V_{inter}$ are the intra-dot and inter-dot Coulomb interactions. From the data shown in Fig. \ref{fig:fig5}, we estimate the value of the inter-dot Coulomb interaction $V^{50(60)}_{inter}\approx 3.4~(2.0)$ meV for $d=50~(60)$ nm, which is given by the DTP separation (for the lowest triplet state) in the limit of large magnetic fields. These numbers are in good agreement with the corresponding expectation values $\left\langle C({\bf r_1},{\bf r_2}) \right\rangle $ of the Coulomb interaction matrix (3.5 and 2.2 meV, respectively) obtained from direct calculations, thereby confirming electron localization and QDs decoupling. Since at zero magnetic field, the DTP separation is equal to $2t+V_{inter}$, we obtain $2t^{50(60)}\approx1.6~(0.7)$ meV which is consistent with the energy differences between the two lowest single-particle levels of 1.9 (0.4) meV. As $V_{intra}\approx 8$ meV is given by the electron addition energy in one QD which is the distance between the "corners" of the linear region where single electron re-localization occurs from one dot to the other in the $N=2$ energy diagram (see Fig. \ref{fig:fig2}), the estimated values of the exchange energy become $J^{50(60)}_{est}\approx 0.6~(0.08)$ meV. These numbers are of the same order as the numerically exact values of 0.24 (0.012) meV, but they both significantly {\it overestimate} the computed data, and therefore, can only be used as a general guideline to gauge the magnitude of the exchange coupling in double QDs. The overestimation is due to the difference between Coulomb energies in the singlet and triplet states that lowers the exchange energy,\cite{Burkard} but which is not taken into account in the simple Hubbard model.

In summary, we computed the stability diagram for model double QD systems populated with up to two electrons in magnetic fields using numerically exact diagonalization of the one- and two-electron Hamiltonian. Two inter-dot separations $d=50$ and 60 nm corresponding to strong and weak inter-dot coupling were considered. We found that in the weak-coupling regime the curvature of the chemical potential $\mu(2)$ is larger than that one of $\mu(1)$ while in the strong-coupling case the situation is reversed. Hence, by analyzing the chemical potential variations caused by external biases and magnetic fields, it is possible to distinguish between strong and weak inter-dot coupling, even if the curvatures are of the same order in both cases. The evolution of the stability diagrams in magnetic fields conforms to the general idea of enhanced electron localization with increasing field strength. The exchange energies extracted from the stability diagrams showed that these values are significantly overestimated when compared with numerically exact data.

This work is supported by DARPA QUIST program through ARO Grant DAAD 19-01-1-0659. The authors thank the Material Computational Center at the University of Illinois through NSF Grant DMR 99-76550. LXZ thanks the Computer Science and Engineering Fellowship Program at the University of Illinois for support.


\clearpage
\begin{center}
FIGURES
\end{center}

\clearpage

\begin{center}
TABLES
\end{center}

\begin{table*}[ht]
\caption{\label{tab:tab1}
Comparison of the magnitude of the second order derivative $\kappa(1)$ and $\kappa(2)$
of the chemical potential curves $\mu(1)$ and $\mu(2)$ near their turning points for various inter-dot separations and magnetic fields. $\kappa^S(2)$ and $\kappa^T(2)$ denote the values for the singlet and the lowest triplet state, respectively.}
\begin{ruledtabular}
\begin{tabular}{ccccc}
 &\multicolumn{2}{c}{$d=50$ nm}&\multicolumn{2}{c}{$d=60$ nm}\\
 B (T)&$\kappa(1)$&$\kappa^S(2)$ [$\kappa^T(2)$]&$\kappa(1)$&$\kappa^S(2)$ [$\kappa^T(2)$] \\ \hline
 $0$&$0.307$&$0.238$ $[0.269]$ &$1.800$&$2.550$ $[2.52]$\\
 $3$&$0.400$&$0.336$ $[0.349]$ &$4.673$&$6.145$ $[7.463]$\\
 $6$&$0.714$&$0.631$ $[0.659]$ &$15.706$&$27.230$ $[37.141]$\\
\end{tabular}
\end{ruledtabular}
\end{table*}


\begin{thebibliography}{99}
\bibitem{note4} The values of the chemical potentials at which the contour plots are drawn are chosen to facilitate data representation. For the investigated magnetic fields, the relative change of the curvature of a specific chemical potential and the DTP separation is less than $5\%$ for different reference values.
\bibitem{note5} The same consideration is obviously valid for the triplet states as well, albeit with a larger quasilinear region.
\end{thebibliography}
\end{document}